\documentclass[12pt]{article}
\usepackage{epsfig}
\newcommand{\bea}{\begin{eqnarray}}
\newcommand{\ena}{\end{eqnarray}}

\begin{document}
\baselineskip=15pt
\begin{titlepage}
\setcounter{page}{0}

\begin{center}
\vspace*{5mm}
{\Large \bf Parametrization of K-essence and Its Kinetic Term}\\
\vspace{15mm}

{\large Hui Li,$^{b}$ \footnote{e-mail address: lihui@itp.ac.cn}
Zong-Kuan Guo$^b$  and
Yuan-Zhong Zhang$^{a,b}$}\\
\vspace{10mm}
\it $^a$CCAST (World Lab.), P.O. Box 8730, Beijing 100080, China
$^b$Institute of Theoretical Physics, Chinese Academy of
   Sciences, P.O. Box 2735, Beijing 100080, China \\

\end{center}

\vspace{20mm} \centerline{\large \bf Abstract} {We construct the
non-canonical kinetic term of a k-essence field directly from the
effective equation of state function $w_k(z)$, which describes the
properties of the dark energy. Adopting the usual parametrizations
of equation of state we numerically reproduce the shape of the
non-canonical kinetic term and discuss some features of the
constructed form of k-essence. } \vspace{2mm}

\begin{flushleft}
\end{flushleft}

\begin{flushleft}
PACS number(s): 98.80.Cq, 98.65.Dx
\end{flushleft}

\end{titlepage}


Recent observations of type Ia supernovae~\cite{AGRSP, wytm},
measurements of the cosmic microwave background~\cite{DNS} and the
galaxy power spectrum~\cite{MT} indicate the existence of the dark
energy $-$ an energy component with a strongly negative equation
of state (EOS) $w_{DE}$ $-$ in addition to matter. For this
purpose, various dynamical scalar field models have been proposed.
Specifically, a reliable model should explain why the present
amount of the dark energy is so small compared with the
fundamental scale (fine-tuning problem) and why it is comparable
with the critical density today (coincidence problem).
Quintessence has been proposed as a candidate for the dark energy
component of the universe that would be responsible of the
currently observed accelerated expansion~\cite{RP}-\cite{PJS}.
Generally speaking, quintessence is a spatially homogeneous field
slow-rolling down its potential. More recently, models based on
scalar fields with non-canonical kinetic energy, dubbed as
k-essence, have emerged~\cite{Picon, wh}. A subclass of models
feature a tracker behavior during radiation domination, and a
cosmological-constant-like behavior shortly after the transition
to matter domination. As long as this transition seems to occur
generically for purely dynamical reasons, these models are claimed
to solve the coincidence problem without fine-tuning. What is
more, lately it has been found that those two theoretical setups
are strongly related to each other and every quintessence model
can be viewed as a k-essence model generated by a kinetic linear
function~\cite{JMA}.

The dark energy is characterized by its equation of state
parameter $w_{DE}$, which is in general a function of redshift $z$
in dark energy models. One could specify a Lagrangian and solve
the scalar field equation for the particular
theory~\cite{Peebles}. Although it may provide a specific form of
$w_{DE}(z)$ in terms of scalar field model parameters which can be
directly constrained through data fitting~\cite{Chen, Watson},
this physically motivated approach clearly limits us to a model by
model test~\cite{Ng}. Apart from the feasibility of quintessence
potential $V(\phi)$ and the EOS $w_{DE}(z)$ reconstruction from
supernova observations~\cite{DHMT, MSAD}, a phenomenological
parametrization of $w_{DE}(z)$ is still necessary to compare
different dark energy models. The key point of this alternative
strategy is to assume a more or less arbitrary ansatz for
$w_{DE}(z)$ which is not necessarily physically justified but
specially designed to give a good fit to the observational data.
This motivation stimulates several parametrization methods to
produce an appropriate parameter space~\cite{SHEM}-\cite{BFGGE}.
Recent study found that it is also workable to drive these
parametrizations to construct the quintessence potential
form~\cite{Guo} and some general features of these constructions
were discussed. In this Letter, we try to carry  out this method
in constructing the k-essence with pressure $p(\varphi,X)$
directly from the dark energy equation of state function
$w_{k}(z)$.  When applying this method to four typical
parametrizations, the shapes of the kinetic term are numerically
obtained and relevant discussions on the resulting construction
are also given.


Restricting our attention to a single field model, the action of
the k-essence generically may be expressed as
\begin{equation}
S_{\varphi }=\int d^{4}x\,\sqrt{-g}\left[ -\frac{R}{6}+p(\varphi
,X)\right] , \label{action}
\end{equation}
where we use units such that $8\pi G/3\equiv1,$ and
\begin{equation}
X=\frac{1}{2}(\nabla \varphi )^{2}.  \label{X}
\end{equation}
The Lagrangian $p$ depends on the specific particle theory model.
In this paper, we consider only factorizable Lagrangians of the
form~\cite{MRL}
\begin{equation}
p=K(\varphi )\widetilde{p}(X),  \label{p}
\end{equation}
where we assume that $K(\varphi )>0.$ To describe the behavior of
the scalar field it is convenient to use a perfect fluid analogy.
The role of the pressure is played by the Lagrangian $p$ itself,
while the energy density is given by \cite{ADM}
\begin{eqnarray}
\varepsilon  &=&K(\varphi
)(2X\widetilde{p}_{,X}(X)-\widetilde{p}(X))
\label{energy} \\
&\equiv &K(\varphi )\tilde{\epsilon}(X),
\end{eqnarray}
where $..._{,X}$ denotes a partial derivative with respect to $X$.
The ratio of pressure to energy density, which we call, for
brevity, the $k$-essence equation-of-state,
\begin{equation}
w_{k}\equiv \frac{p}{\varepsilon
}=\frac{\widetilde{p}}{\tilde{\epsilon}}=
\frac{\widetilde{p}}{2X\widetilde{p}_{,X}-\widetilde{p}}\,,
\label{w-k}
\end{equation}
does not depend on the function $K(\varphi )$.

 We will consider a spatially flat FRW universe which is dominated by the
non-relativistic matter and a spatially homogeneous scalar field
$\varphi$. The Friedmann equation can be written as
\begin{equation}
H^{2}\,=\rho _{m}+\varepsilon , \label{friedmann}
\end{equation}
where $\rho_m$ is the matter density. The evolution of k-essence
field is governed by the equation of motion
\begin{equation}
\label{emp} \dot{\varepsilon}+3H(\varepsilon+p)=0,
\end{equation}
which yields
\begin{eqnarray}
\varepsilon (z) &=& \varepsilon_{0}
 \,\exp \left[3\int_{0}^{z}(1+w_k)d\ln (1+z)\right]
 \nonumber \\
 & \equiv & \varepsilon_{0}\,E(z),
\label{rhoz}
\end{eqnarray}
where $z$ is the redshift which is given by $1+z = a_0/a$ and
subscript $0$ denotes the value of a quantity at the redshift
$z=0$ (present). As the constant $\varepsilon_{0}$ is of no
importance for further discussion,we will safely set it to $1$ and
then regard the energy density $\varepsilon(z)$ as a dimensionless
quantity.

So we get $\varepsilon(z)$ as the function of redshift $z$, and by
means of the EOS of k-essence $\widetilde{p}(z)$ may be well
prepared as follows,
\begin{equation}
\label{pz1}\widetilde{p}(z)=\frac{w_{k}(z)E(z)}{K(\varphi)}.
\end{equation}
Once the form of the potential function $K(\varphi)$ is specified,
$\varphi(z)$ supplied by the Eq.~(\ref{vphizz}) will describe the
behavior of $\widetilde{p}(z)$. With the help of
Eq.~(\ref{energy}) and Eq.~(\ref{w-k}), $X$ can also be implicitly
expressed as the function of $z$:
\begin{equation}
\label{Xz} \varepsilon(z)\,=\,K(\varphi) \left[2 X
\frac{d\widetilde{p}(z)}{dz}\frac{dz}{dX}-\widetilde{p}(z)\right].
\end{equation}
Therefore, the redshift relates two functions $p(z)$ and $X(z)$,
and the reconstruction may be implemented. Using
$\rho_m=\rho_{m0}(1+z)^3$ and Eq.~(\ref{rhoz}), the Friedmann Eq.~
(\ref{friedmann}) becomes
\begin{equation}
\label{friedmannzz} H(z)=H_0\left[\Omega_{m0}(1+z)^3+\Omega_{k0}
 E(z)\right]^{1/2},
\end{equation}
where $\Omega_{m0} \equiv \rho_{m0}/H_0^2$ and $\Omega_{k0} \equiv
\rho_{k0}/H_0^2$. Due to the homogeneity of k-essence
$X=\dot\varphi^2/2$ and therefore,
\begin{equation}
X(z)=\frac{(\varphi'(z))^2(1+z)^2 H(z)^2}{2} \label{xzz}\,,
\end{equation}
where the dot represents derivative with respect to the physical
time $t$ and the prime derivative with respect to the redshift
$z$. As a consequence,
$\dot{\varphi}\equiv\varphi'(z)dz/dt=\pm\sqrt{2 X}$. Because
$\tilde p(x)$ is only dependent on the variable $X$, the sign of
$\dot\varphi$ is insignificance for the construction process
below.

For definiteness we will choose
$K(\varphi)=\varphi^{-2}$~\cite{MRL}. Since $dz/dt=-(1+z)H(z)$ and
$H(z)$ is given by Eq.~(\ref{friedmann}), after straightforward
calculations the differential equation for $\varphi(z)$ will be
written as the Bernoulli type

\begin{equation}
v'(z)+\left[\frac{1}{1+z}+\frac{H'(z)}{H(z)}-\frac{w'_{k}}{1+w_{k}}-\frac{w_{k}E'(z)}{(1+w_{k})E(z)}\right]v(z)-\frac{w_{k}-1}{w_{k}+1}v(z)^2=0,
\label{vphizz}
\end{equation}
where $v(z)\equiv\varphi'/\varphi$.

After identifying $\varphi$ with the dimensionless quantity
$\tilde{\varphi} \equiv \varphi / M_{pl}$, the construction
equations are of the same form as Eq.~(\ref{pz1}) and
Eq.~(\ref{xzz}), which relate the non-canonical kinetic term
$\tilde{p}(X)$ to the equation of state function $w_{k}(z)$. Given
an effective equation of state function $w_{k}(z)$ ,
Eq.~(\ref{vphizz}) will make these two construction equations
(\ref{pz1}) and (\ref{xzz}) simulate the non-canonical kinetic
energy $\tilde{p}(X)$.

Our method relates directly the kinetic terms of k-essence to the
equation of state function, and so enables us to construct easily
the former without assuming its form. The dark energy properties
are well described by the effective equation of state parameter
$w_k(z)$ which in general depends on the redshift $z$.

Following the previous work~\cite{Guo}, we consider the
construction process with the following parametrization
methods~\cite{SHEM}-\cite{BFGGE}: a constant equation of state
parameter and the other three two-parameter parametrizations,

{\bf Case 1}: $w_k=w_0$ (Ref.~\cite{SHEM})

{\bf Case 2}: $w_k=w_0+w_1 z$ (Ref.~\cite{ARCDH})

{\bf Case 3}: $w_k=w_0+w_1 z/(1+z)$ (Refs.~\cite{CP,EVL,TP})

\begin{figure}
\begin{center}
\includegraphics[width=11cm]{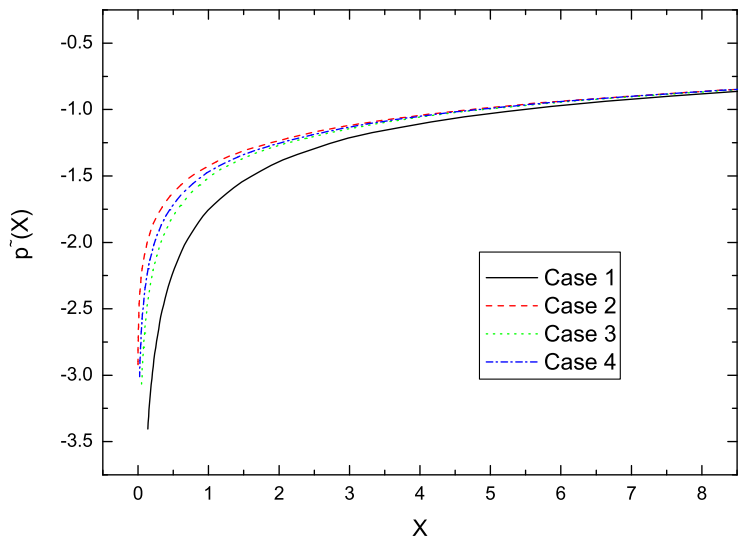}
\caption{Constructed kinetic term $\tilde{p}(X)$ in the k-essence
model.} \label{fig:px}
\end{center}
\end{figure}

{\bf Case 4}: $w_k=w_0+w_1\ln(1+z)$ (Ref.~\cite{BFGGE})

we choose $w_0=-0.7$, $w_1=-0.2$ and $r_0=3/7$ for the specific
construction~\cite{BARGER}. This is quite different from the
quintessence case as $dw_k/dz>0$ is generally required for the
former~\cite{Picon, YWang}. This noticeable feature would enable
supernova data to distinguish between the two theories, although
some ambiguity is sticky to deal with. Just for definiteness we
have chosen the initial values of the k-essence field
$\varphi_0=1.0$ and $v_0=0.1$ at $z=0$. This degree of freedom of
initial value settings will be easily eliminated after appropriate
field rescaling and  have no essential influence on the general
form of $\tilde{p}(X)$. As a matter of fact, the change of the
initial value $\varphi_0$ or $v_0$ just lifts the corresponding
curves vertically and keeps their shapes intact.The numerical
results are plotted as follows. Fig.~\ref{fig:px} shows the
constructed non-canonical kinetic term $\tilde{p}(X)$. At low
redshift region($z<1$, corresponding to $X>4$), all these models
share the same asymptotical behavior, but begin to deviate from
this around redshift $z>1$.

We should note that the parametrizations we adopted here might not
be of particular significance but just for concreteness. Since
there is little theoretical guidance as to the nature of the dark
energy, the phenomenological parametrization of $w_k(z)$ should be
designed as generally as possible unless no observations can
effectively unravel such excess complication. It has been shown
that more than two additional parameters is not viable for a
general fit with next generation data but two parameter
approximations in current use are reasonable and realistic for the
near future observations~\cite{YWang, Linder, Caldwell}. On the
other hand, due to the lack of data and the large statistical
error bars, a wide range of dark energy models seem to have ranges
of parameters that fit the existing data equally well and no
particular model stands out as being observationally preferred. As
a consequence, any such parametrizations made in potential
reconstruction methods may be too restrictive since many different
potential energy functions are conceivable and many of them may
give results degenerate with each other. While it would still be
difficult to break this observational degeneracy, the challenging
goal of making the fundamental physics distinction of the sign of
the EOS time variation is achievable in the next generation of
experiments~\cite{Linder, Caldwell}. One of the direct rewards
relevant to the two parameter approximation then would be that the
usual form of k-essence with $dw_k/dz>0$ could be knocked down.
Anyhow, although we have concentrated specially on the k-essence
form in Ref.~\cite{MRL}, it is obvious that the techniques
exploited here may be applied to more general k-essence models as
well.


In conclusion, we have used the scheme of constructing the form of
dark energy developed in~\cite{Guo} directly from the effective
equation of state function $w_k(z)$. Then we have considered four
parametrizations of equation of state parameter and showed that
the constructed k-essence model may also be constructed without
further dynamical information of the k-essence field. Current
popular parametrizations we have referred to are directly related
to the data of type Ia supernovae; therefore, our approach will be
a useful tool to reexamine the k-essence models, and the two
parameters extracted from the SNe Ia observations will not only
help us select the suitable k-essence form or rule out the usual
form of k-essence models with great confidence in the future, but
also may lead to a strong constraint and convenient guidance for
the fundamental theory of the perplexing dark energy. The future
Supernova/Acceleration Probe with high-redshift
observations~\cite{SNAP}, in combination with the Planck CMB
observation~\cite{PLANCK}, will be able to determine the
parameters in the dark energy parametrization to high precision.
By precision mapping of the recent expansion history, we hope to
learn more about the essence of the dark energy and get a deeper
understanding of the dynamics of the universe.

\section*{Acknowledgements}
This project was in part supported by National Basic Research
Program of China under Grant No. 2003CB716300 and by NNSFC under
Grant No. 90403032.

\end{document}